# Did Poincaré explore the inertial mass-energy equivalence?

Galina Weinstein[*]

Einstein was the first to explore the inertial mass-energy equivalence. In 1905 Einstein showed that a change in energy is associated with a change in inertial mass equal to the change in energy divided by $c^2$. In 1900 Poincaré considered a device creating and emitting electromagnetic waves. The device emits energy in all directions. As a result of the energy being emitted, it recoils. No motion of any other material body compensates for the recoil at that moment. Poincaré found that as a result of the recoil of the oscillator, in the moving system, the oscillator generating the electromagnetic energy suffers an "apparent complementary force". In addition, in order to demonstrate the non-violation of the theorem of the motion of the centre of gravity, Poincaré needed an arbitrary convention, the "fictitious fluid". Einstein demonstrated that if the inertial mass $E/c^2$ is associated with the energy E, and on assuming the inseparability of the theorem of the conservation of mass and that of energy, then – at least as a first approximation – the theorem of the conservation of the motion of the centre of gravity is also valid for all systems in which electromagnetic processes take place. Before 1905 (and also afterwards) Poincaré did not explore the inertial mass-energy equivalence. In 1908 Einstein wrote the German physicist Johannes Stark, "I was a little surprised to see that you did not acknowledge my priority regarding the relationship between inertial mass and energy".

Hendrik Antoon Lorentz's theory of the electron violated the principle of action and reaction, but Henri Poincaré believed he could mend this violation.[1] Perhaps because Lorentz was so successful in deriving Fresnel's dragging coefficient in his theory, Poincaré also believed that, one could find an explanation that would yield conservation of momentum (for ether and matter) in Lorentz's theory.

Poinacré devoted his talk "La théorie de Lorentz et le principe de réaction" (The Theory of Lorentz and the Principle of Reaction) for solving this problem. He presented the talk in the celebrations (Festschrift) of the 25th anniversary of Lorentz's doctorate in Leiden on December 11, 1900, while Lorentz was sitting in the audience. The talk was published later as a paper in the *Archives neérlandaises*.[2]

## 1. Poincaré 1900 – The "Hertzian Oscillator"

Poincaré considered a "Hertzian Oscillator", a device like the one used by Heinrich Hertz to create and emit electromagnetic waves. The oscillator emits energy in all directions. Poincaré provided it with a parabolic mirror, as Hertz himself had done with his oscillator, in order for the energy to be sent in a single direction. As a result of the energy being emitted, the device recoils. Although the recoil is very feeble, it exists and might be verified experimentally.

---

When one thinks of the recoil the oscillator undergoes: no motion of any other material body compensates for the recoil at that moment. If the radiation later encounters some material body, the compensation will be effected on the condition that no energy has been lost on the way; in this manner the receiving body becomes a perfect absorbent and absorbs all the energy. It is very likely that the compensation will only be partial, because not all the radiation will reach the receiving body; and even if all of it had reached that body, the compensation *will not be simultaneous*. The compensation is effected by the radiation pressure on the bodies it encounters; if it never encounters any material body, the compensation will never be effected.[3]

Poincaré found that one had to assert that, there existed compensation, an "Apparent Complementary Force".[4] The oscillator, being in "real motion" with respect to the ether, radiates energy density of amount J. For an observer carried along with a system moving relative to the oscillator's system, being in an "apparent motion," the energy is $J(1 – v/c)$, where v is the apparent velocity of the system relative to the oscillator. The "real recoil" measured in the oscillator's system is $J/c$. The "apparent recoil" measured in the system moving relative to the oscillator's is, $J/c (1 – v/c)$ or, $J/c – Jv/c^2$. Therefore, in order to obtain the real recoil, one has to add to the apparent recoil "une force complémentaire apparente", $– Jv/c^2$.

Poincaré added:[5] "(I put the sign – because the recoil, as its name indicates, takes place in the negative direction)." As a result of the recoil of the oscillator, in a system moving relative to the oscillator's system, the oscillator generating the electromagnetic energy suffers an apparent complementary force. In the system of the oscillator, there is no complementary force. The principle of reaction is thus applicable to matter alone because of the existence of the apparent complementary force, even though the principle is in fact also valid for the ether. Poincaré did not explore here a relation between inertial mass and energy.

**2. Einstein 1905 – The Mass of a Body is a Measure of its Energy Content**

On 27 of September, 1905, Einstein published in the *Annalen Der Physik* his paper, "Does the Inertia of a body depend upon its energy content?"

A body at rest in an inertial reference frame *(x, y, z)* sends out – in the line making an angle *φ* with the *x* axis – plane light to both directions with a total energy *L*: each direction it sends waves of energy *L/2*. i.e., since the emissions are diametrically opposed the body loses energy but not momentum, so it remains at rest in the original inertial system *(x, y, z)*.

Einstein defines the following terms: $E_0$ – the initial energy of the body at rest with respect to *(x, y, z)* before, and $E_1$ – after emitting the waves.

$H_0$ – the energy of the body at rest with respect to *(ξ, η, ζ)* before, and $H_1$ – after emitting the waves.

According to the principle of conservation of energy (applicable to both systems as a result of the principle of relativity), Einstein arrives at the solution:[6]

With respect to $(x, y, z)$: $E_0 = E_1 + [L/2 + L/2]$.

In the relativity paper Einstein used the Maxwell-Hertz equations for empty space, together with Maxwell's expression for the electromagnetic energy, and the principle of relativity (and the principle of the constancy of the velocity of light) and deduced the transformation for the energy of light complex in §8:[7]

$l^* = l\,(1 - v/c\,\cos\varphi) / \sqrt{(1 - v^2/c^2)}$

This is the energy of light waves with a normal and an angle $\varphi$ with respect to the x axis measured in the system $(\xi, \eta, \zeta)$, which is moving with respect to the system $(x, y, z)$. With respect to $(\xi, \eta, \zeta)$ using the transformation for the energy written above:

$H_0 = H_1 + [L/2 \cdot (1 - \cos\varphi\,v/c)/\sqrt{(1 - v^2/c^2)}] + [L/2 \cdot (1 + \cos\varphi\,v/c)/\sqrt{(1 - v^2/c^2)}]$

$= H_1 + L/\sqrt{(1 - v^2/c^2)}$,

Subtracting we get:

(1) $(H_0 - E_0) - (H_1 - E_1) = L\,[(1/\sqrt{1 - v^2/c^2}) - 1]$.

Einstein then writes,[8]

"H and E are energy values of the same body, referred to two systems of coordinates in motion relative to each other, the body is at rest in one of the two systems (system $(x, y, z)$). *It is therefore clear* that the difference H − E can differ from the kinetic energy K of the body, with respect to the other system $(\xi, \eta, \zeta)$, only by an additive constant C, which depends on the choice of the arbitrary additive constants of the energies H and E".

We may therefore set,

(2)   $H_0 - E_0 = K_0 + C$

   $H_1 - E_1 = K_1 + C$,

since C does not change during the emission of light.

Substituting equation (2) into equation (1) gives:

(3) $K_0 - K_1 = L\{[1/\sqrt{(1 - v^2/c^2)}] - 1\}$

(before and after the emission).

After this expression Einstein writes, "The kinetic energy of the body with respect to $(\xi, \eta, \zeta)$ decreases as a result of the emission of light, and by an amount that is independent of the properties of the body".[9] Einstein immediately notes afterwards, that, the difference of the two kinetic energies (3) depends on the velocity in the same way as the kinetic energy of the electron, the expression that he had obtained in §10.

Finally, Einstein draws his conclusion from equation (3) by passing to the Newtonian limit.[10] Neglecting magnitudes of the fourth and higher order, he arrives at:

(4) $K_0 - K_1 = \frac{1}{2} L v^2/c^2$.

Einstein concluded, "If a body emits the energy $L$ in the form of radiation, its mass diminishes by $L/c^2$ ".[11]

Einstein ended his paper by saying,[12]

"It is inessential that the energy withdrawn from a body turns straight into radiation energy, so we are led to the general conclusion:

The mass of a body is a measure of its energy content; if the energy changes by $L$, the mass changes in the same sense by $L/9 \cdot 10^{20}$ if the energy is measured in ergs and the mass in grams".

In 1905 Einstein was only able to show that a change in energy is associated with a change in inertial mass equal to the change in energy divided by $c^2$. Einstein returned to the relation between inertial mass and energy in 1906 – as shown below, and in 1907, giving more general arguments for their complete equivalence.[13]

**3. Poincaré 1900 – The Fictitious Fluid**

In the 1900 Lorentz Festschrift paper Poincaré considered a small conductor, charged positively and surrounded by ether. An electromagnetic wave passes through the ether and strikes the conductor. The wave had originated from a source from which the waves were completely detached when they hit the conductor. At the moment the waves reached the conductor, the electric force resulting from the perturbation would act on the charge, causing a pondermotive force to act on the conductor. In accordance with the principle of reaction, this force would not be balanced by any other force acting on ponderable matter because all other ponderable bodies are supposed to be far removed, well beyond the scope of the perturbed ether.

It appears that the solution could be found in the ether: if we suggest a detailed mechanistic model for the ether, then the ether would cause the action on the ponderable matter and this would solve the problem. Joseph Larmor, in accordance with the British tradition, had thought of this solution. However, Poincaré objected to mechanistic models of the ether. He explained that if a mechanical model of electromagnetic ether does indeed exist, then an infinite number of such models must exist as well. Therefore, it was useless to invoke a mechanical explanation for ether when one can find infinite mechanical explanations to the phenomena. Poincaré wrote in his lectures on electricity and optics (*Électricité et optique*),[14]

*"If, then a phenomenon involves a complete mechanical explanation, it will involve an infinite of others that will equally well report all the particularities revealed by experience".*

In a 1921 manuscript, "Fundamental Ideas and methods of the Theory of Relativity, Presented in Their Development", Einstein explained Poincaré's above reasoning,[15]

"Maxwell himself still clung to the conception that all physical events have to be interpreted in terms of mechanics. But his efforts, and those of other important theoreticians, to devise a mechanical model of electromagnetic phenomena in the ether did not meet with success. Poincaré pointed out that even if the construction of such a picture were accomplished, it would not be a decisive success because such a picture would only be one in a an infinite number of possible ones which, in principle, would be equally justified".

In his 1900 Lorentz Festschrift paper Poincaré assumed that he could mathematically reestablish the equality of action and reaction in the theory of Lorentz for the phenomena of radiation pressure.[16] Poincaré considered electrons inside volume V, on which other far-removed electrons, positioned outside the volume, acted by means of electromagnetic energy transmitted by the ether. Consider an element of small volume dV containing electrons with charge ρdV. The resultant of all the forces that are exerted on the electrons is:[17]

(A) $\mathbf{F} = \int \rho dV (\mathbf{E} + \mathbf{v}/c \times \mathbf{B})$.

where $f$ is the Lorentz force density, $\mathbf{B}$ is the magnetic field, $\mathbf{E}$ is the electric field, ρ is the electric charge density, $\mathbf{v}$ is the velocity of the electrons, and c is the velocity of light. ρ$\mathbf{E}$dV represents the action of the electric field, and (ρ$\mathbf{v}$/c × $\mathbf{B}$)dV is the action of the magnetic field.

Poincaré transformed equation (A) into an expression involving the magnetic and electrostatic Maxwell pressures [which he wrote as equation (1)] that act on a surface enclosing the volume of integration, and,[18]

(2) $\mathbf{F} = -d/dt \cdot 1/4\pi c \int (\mathbf{E} \times \mathbf{B}) dV$ + Maxwell's electrostatic and magnetic pressures.

Without the presence of the first term, the force acting on the electrons would result only from Maxwell's pressures. These pressures disappear when the surface recedes to infinity, because the field there vanishes. Still the total resultant of the actions on the electrons does not vanish,[19]

(C) $\mathbf{F} = -d/dt \cdot 1/4\pi c \int (\mathbf{E} \times \mathbf{B}) dV$.

This precisely indicates that Lorentz's theory did not satisfy the principle of reaction. The sum of the momenta of the ensemble of electrons does not remain constant.

Consider,

(D) d/dt $\sum m\mathbf{v} = \mathbf{F}$.

$\mathbf{F}$ = the electromagnetic forces and the non-electromagnetic collisions (actions due to other electrons); m is the mass of the electrons, and $\mathbf{v}$ is their velocities.

If the principle of reaction is valid for matter (the electrons) alone, one finds in this system that the conservation of momentum is satisfied when there is no electromagnetic field $\mathbf{F}=0$. Thus, d/dt $\sum m\mathbf{v} = 0$, and, [20]

(E) $\sum m\mathbf{v}$ = const.

But when, $\mathbf{F} \neq 0$, according to (C) we obtain,

d/dt $\sum m\mathbf{v} = -$d/dt $\cdot$ 1/4πc $\int (\mathbf{E} \times \mathbf{B})dV$.

The following conservation of momentum law is obtained, [21]

(F) $\sum m\mathbf{v}$ + 1/4πc $\int (\mathbf{E} \times \mathbf{B})dV$ = const.

Poincaré indicated that, in order for the above equation to be valid, one has to regard electromagnetic energy as a fluid carrying with it a total momentum,

(G) $-$1/4πc $\int (\mathbf{E} \times \mathbf{B})dV$.

If the momentum is the momentum of the electromagnetic energy, which is localized in the ether, and considered as a mass animated with a certain velocity, according to the ideas of Maxwell, one arrives at the existence of radiation pressure. One then has to say that the principle of reaction is valid for matter plus the ether.[22] However, Poincaré believed that any experiment whatsoever conducted for the purpose of confirming this conclusion would always give the result that the principle of reaction was applicable to matter alone.[23]

Poincaré defined the fluid as a "Fictitious Fluid" ("fluide fictif"), and wrote (G) as:

$\int J/c^2 \cdot \mathbf{u}dV$,

and $\int J/c^2 \cdot \mathbf{u}$ is the quantity of fictitious fluid passing through a unit surface per unit time: $J/c^2$ is the mass density of the fictitious fluid, and $\mathbf{u}$ is its velocity.[24]

Poincaré then tried to solve the problem by writing, [25]

(H) $\int J/c^2 \cdot \mathbf{u}dV$ = 1/4πc $\int (\mathbf{E} \times \mathbf{B})dV$,

where the energy density is the Poynting equation: $J = (E^2 + B^2)/8\pi$.

Inserting (H) into equation (F) gives, [26]

(4) $\sum m\mathbf{v} + \int \mathbf{J}/c^2 \cdot \mathbf{u}dV$ = const.

In ordinary mechanics, from the constancy of the momentum, it has been concluded that the motion of the centre of gravity is rectilinear and uniform. However, from the above equation it cannot be concluded that the centre of gravity of the system is formed of matter and the fluid is in a rectilinear and uniform motion, because the fictitious fluid is not indestructible. The fluid is destroyed/created per unit volume dV and time by a quantity: $1/c^2 \int \rho(\mathbf{v} \cdot \mathbf{E})dV$.[27]

The position of the centre of gravity of the fictitious fluid is: $\int \mathbf{J}/c^2 \cdot \mathbf{r}dV$ and of matter it is: $\sum m\mathbf{r}$.[28]

Using (4) Poincaré obtained, [29]

(3) $d/dt \sum m\mathbf{r} + 1/c^2 \int \rho(\mathbf{v} \cdot \mathbf{E})dV$ = const.

**4. Lorentz's Response to Poincaré's 1900 Paper**

On January 20, 1901 Lorentz wrote a letter to Poincaré and responded to the latter's 1900 Lorentz Festschrift paper.[30] Lorentz was especially troubled by Poincaré's equation (F).[31] Lorentz seemed not to accept Poincaré's idea that, in order for this equation to be valid, one has to regard electromagnetic energy as a fictitious fluid carrying with it a total momentum.[32]

Lorentz told Poincaré: it is true that when a body acquires a certain amount of motion, then the mind is satisfied only when we can identify a simultaneous change in another body. "But I think we can also be satisfied when the change is not simultaneous, see even the product of motion you have deduced form the beautiful formula" (F).[33]

Lorentz then told Poincaré, [34]

"I think we can simply consider […$1/4\pi c \int (\mathbf{E} \times \mathbf{B})dV$]

to be quantities dependant on the state of the ether, which are 'equivalent' to the momentum".

And Lorentz thought, "we could just be satisfied with this".[35] Of course Poincaré could not accept this solution. However, Lorentz added, "I must confess that I cannot change my theory in the way you report that the difficulty disappears".[36]

Lorentz explained to Poincaré, "It seems unlikely that we will gain success. I rather think (and these results are also implicit in your comments) that the violation of the principle of reaction is essential in all theories that are able to explain Fizeau's experiment".[37] Lorentz realized that he had to choose between Fizeau's experiment and the principle of action and reaction, and he told Poincaré, "As to the principle of reaction, it does not seem to be a fundamental principle of physics".[38]

In 1908 Poincaré chose Fizeau's experiment and he gave up the principle of reaction.[39] However, in his lessons at the Ecole Supérieure of 1911-1912 "The Dynamics of the Electron", posthumously published from the notes of his students and a list of equations left by Poincaré, he returned to the problem and to the "fluide fictif".[40]

**5. Einstein 1906 – inseparability of theorem of conservation of mass and of energy**

In his 1906 paper, "Das Prinzip von der Erhaltung der Schwerpunktsbewegung und die Trägheit der Energie" (The Principle of Conservation of Motion of the Centre of Gravity and the Inertia of Energy"), Einstein solved Poincaré's problem.

Einstein mentioned Poincaré's 1900 paper in this regard. He wrote that the simple formal considerations he had used were already contained in Poincaré's work, but he had preferred not to base himself on that work for the sake of clarity.[41]

From the Maxwell-Lorentz equations Einstein obtained the following equation,[42]

(5) $1/c^2 \int \rho \mathbf{r}(\mathbf{v} \cdot \mathbf{E})dV + d/dt \int \mathbf{J}/c^2 \cdot \mathbf{r}dV - 1/4\pi c \int (\mathbf{E} \times \mathbf{B})dV = 0$.

Only if there is no destruction or creation of electromagnetic energy anywhere, is the centre of gravity of a system formed of matter and the electromagnetic energy will have a rectilinear and uniform motion and one obtains Poincaré's equation (3).

Einstein first rewrote equation (5) in the following way,[43]

(6) $\sum (\mathbf{r} \cdot dm/dt) + d/dt \int 1/c^2 \int \rho \mathbf{r}dV - 1/4\pi c \int (\mathbf{E} \times \mathbf{B})dV = 0$.

Poincaré did not write this equation. The second term meant the proportionality of mass to the energy divided by $c^2$. As a consequence, in 1900 Poincaré did not associate an inertial mass $E/c^2$ to the energy E. Inserting equation (6) into Poincaré's equation (F) leads to Poincaré's equation (3).

Einstein demonstrated that if the inertial mass $E/c^2$ is associated with the energy E, and on assuming the inseparability of the theorem of the conservation of mass and that of energy, then – at least as a first approximation – the theorem of the conservation of the motion of the centre of gravity is also valid for all systems in which electromagnetic processes take place.[44] From equation (6) Einstein obtained Poincaré's equation (3).[45]

In order to demonstrate the non-violation of the theorem of the motion of the centre of gravity from his equation (3), Poincaré needed an arbitrary convention. Einstein demonstrated this from (3) by using the assumption that the theorem of the constancy of mass is a special case of the principle of energy. From equation (6) Einstein obtained,[46]

$X = (\sum(m\mathbf{r}) + 1/c^2 \int \rho \mathbf{r}dV)/(\sum m + 1/c^2 \int \rho dV)$,

where X is the x coordinate of the centre of gravity of the material masses and of the "mass energy" of the electromagnetic field. The denominator is independent of time according to the mass-energy principle, and can be equated to a constant.

Therefore, [47]

(7)  $dX/dt = $ const.

This equation represents the law of conservation of the motion of the centre of gravity.

Einstein was the first to explore the inertial mass-energy equivalence. Dukas and Hoffmann report that on 17 February 1908, a somewhat aggravated Einstein in the Patent Office in Bern wrote a postcard to the German physicist Johannes Stark, who was later to receive the Nobel Prize: "I was a little surprised to see that you did not acknowledge my priority regarding the relationship between inertial mass and energy".[48] Stark answered at that time warmly and regretted. Einstein indeed deserved the priority for the relationship between inertial mass and energy.

*I wish to thank Prof. John Stachel from the Center for Einstein Studies in Boston University for sitting with me for many hours discussing special relativity and its history. Almost every day, John came with notes on my draft manuscript, directed me to books in his Einstein collection, and gave me copies of his papers on Einstein, which I read with great interest*

$$X = \int \rho d\tau \left[\eta\gamma - \zeta\beta \frac{4\pi f}{K_0}\right].$$

The integration is performed over the volume element dτ, and ξ, η, ζ are the components of the velocity of the electron, and the velocity of light is 1.

Poincaré wrote,

$$\rho\eta = -\frac{dg}{dt} + \frac{1}{4\pi}\left(\frac{d\alpha}{dz} - \frac{d\gamma}{dx}\right);$$

$$\rho\zeta = -\frac{dh}{dt} + \frac{1}{4\pi}\left(\frac{d\beta}{dx} - \frac{d\alpha}{dy}\right);$$

$$\rho = \sum \frac{df}{dx}.$$

[18] Poincaré, 1900, pp. 254-255. Poincaré obtained the expression for the magnetic Maxwell pressure from (A),

(1) $X_2 + X_3 =$

$$\int \frac{d\omega}{8\pi}[l(\alpha^2 - \beta^2 - \gamma^2) + 2m\alpha\beta + 2n\alpha\gamma].$$

X = $X_1 + X_2 + X_3 + X_4$, and l, m, n are the direction cosines.

And the electrostatic Maxwell pressure,

$$X'_4 - Y = \int \frac{2\pi d\omega}{K_0}[l(f^2 - g^2 - h^2) + 2mfg + 2nfh].$$

And equation (2) in Poincaré's notation is,

(2) $X = \dfrac{d}{dt}\displaystyle\int d\tau\,(\beta h - \gamma g) + (X_2 + X_3) + (X'_4 - Y).$

[19] Poincaré, 1900, p. 255. In Poincaré's notation,

$$X = \frac{d}{dt}\int d\tau\,(\beta h - \gamma g).$$

[20] Poincaré, 1900, p. 255. And in Poincaré's notation,
$\sum mV_x = $ const, $\sum mV_y = $ const, $\sum mV_z = $ const.

[21] Poincaré, 1900, p. 255. In Poincaré's notation,

$$\sum MV_x + \int d\tau\,(\gamma g - \beta h) = const.$$

$$\sum MV_y + \int d\tau\,(\alpha h - \gamma f) = const.$$

$$\sum MV_z + \int d\tau\,(\beta f - \alpha g) = const.$$

[22] Poincaré, 1900, p. 278.

[23] Poincaré, Henri, "A propos de la théorie de M. Larmor", *L'eclairage électrique* 3, pp. 5-13; pp. 289-295; *L'eclairage électrique* 5, 1895, pp. 5-14, pp. 385-392, p. 392.

[24] Poincaré, 1900, p. 256. In Poincaré's notation, the fictitious fluid had density $K_0J$, and the components of the velocity of the fluid are $U_x, U_y, U_z$; hence, the quantity of fictitious fluid passing through a unit surface per unit time is, $K_0JU_x, K_0JU_y, K_0JU_z$.

[25] Poincaré, 1900, p. 256. In Poincaré's notation, Since,
$K_0JU_x = \gamma g - \beta h$
$K_0JU_y = \alpha h - \gamma f$
$K_0JU_z = \beta f - \alpha y$
and the Poynting equation is,
$$J = \frac{1}{8\pi}\sum \alpha^2 + \frac{2\pi}{K_0}\sum f^2.$$

[26] Poincaré, 1900, p. 257. In Poincaré's notation,

$$(4) \begin{aligned} \sum MV_x + \int K_0JU_x d\tau = const. \\ \sum MV_y + \int K_0JU_y d\tau = const. \\ \sum MV_z + \int K_0JU_z d\tau = const. \end{aligned}$$

[27] Poincaré, 1900, p. 256. In Poincaré's notation the fluid is destroyed/created per unit volume $d\tau$ and time by the following quantity,
$$\frac{4\pi}{K_0}\rho d\tau \sum f\xi.$$

[28] Poincaré, 1900, p. 258.

[29] Poincaré, 1900, p. 258. In Poincaré's notation, using equation (4), the position of the centre of gravity of matter is,
$$\frac{d}{dt}(M_0X_0) = \sum MV_x.$$
And the position of the centre of gravity of the fictitious fluid,
$$K_0 \int xJd\tau = M_1X_1.$$
Combining both these results gives equation (3),
$$(3)\ \frac{d}{dt}(M_2X_2) = constant - 4\pi \int \rho x d\tau \sum f\xi,$$

where, $M_2 = M_0 + M_1$, and $M_2X_2 = M_0X_0 + M_1X_1$.
$M_0$ – the total mass of matter. $X_0$ – the X component of its centre of gravity.
$M_1$ – the total mass of fictitious fluid. $X_1$ – the X component of its centre of gravity.

[30] Lorentz to Poincaré, 20 January, 1901, letter 38.1, Walter, Scott, Bolmont Étienne, and Coret, André, *La Correspondance entre Henri Poincaré et les physiciens, chimistes et ingénieurs*, 2000, Berlin: Birkhäuser.

[31] Poincaré, 1900, p. 255.

[32] Poincaré, 1900, p. 278.

[33] Lorentz to Poincaré, 20 January, 1901, letter 38.1, Walter, Bolmont, and Coret (ed), 2000, p. 253.

[34] Lorentz to Poincaré, 20 January, 1901, letter 38.1, Walter, Bolmont, and Coret (ed), 2000, p. 253. Lorentz used Poincaré's original notation:

$$\int d\tau\, (\gamma g - \beta h), \int d\tau\, (\alpha h - \gamma j), \int d\tau\, (\beta j - \alpha g).$$

[35] Lorentz to Poincaré, 20 January, 1901, letter 38.1, Walter, Bolmont, and Coret (ed), 2000, p. 253.

[36] Lorentz to Poincaré, 20 January, 1901, letter 38.1, Walter, Bolmont, and Coret (ed), 2000, p. 253.

[37] Lorentz to Poincaré, 20 January, 1901, letter 38.1, Walter, Bolmont, and Coret (ed), 2000, p. 253.

[38] Lorentz to Poincaré, 20 January, 1901, letter 38.1, Walter, Bolmont, and Coret (ed), 2000, p. 253.

[39] Poincaré, Henri, "La dynamique de l'électron", *Revue générale des sciences pures et appliqués* 19, 1908, pp. 386-402 in Poincaré, Henri, *Oeuvres 1916-1965,* Paris: Gauthier-Villars, 11 Vol. 9, pp. 551-586; p. 570.

[40] Poincaré, Henri, "La dynamique de l'electron", *Supplément aux Annales des Postes télégraphes et téléphones*, ed by A. Dumas, March 1913, pp. 26-32.

[41] Einstein, Albert, "Das Prinzip von der Erhaltung der Schwerpunktsbewegung und die Trägheit der Energie", *Annalen der Physik* 20, 1906, pp. 627-633; p. 627.

[42] Einstein, 1906, p. 630. Einstein wrote, "If we also assign to the electromagnetic field a mass density ($\rho_e$), which differs by a factor $1/c^2$ from the energy density, then the second term of the equation takes the form $c^2 \frac{d}{dt}\{\int x\rho_e d\tau\}$".

And "The first term of this equation represents the energy supplied by the electromagnetic field to the bodies m₁,…mₙ. According to our hypothesis on the dependence of the masses on energy, the first term" is:

$$c^2 \sum x_\nu \frac{dm_\nu}{dt},$$

since we assume according to the above that the individual material points $m_\nu$ change their energy, and thereby also their mass, *only by* taking up electromagnetic energy".

[43] Einstein, 1906, p. 631. In Einstein's notation,

$$\sum \left[ x_\nu \frac{dm_\nu}{dt} \right] + \frac{d}{dt}\left\{ \int x\rho_e d\tau \right\} - \frac{1}{4\pi c} J = 0$$

[44] Einstein, 1906, p. 633.

[45] Einstein, 1906, p. 632. In Einstein's notation,

$$\frac{d}{dt}\left[ \sum (m_\nu x_\nu) + \int x\rho_e d\tau \right] = \text{const.}$$

[46] Einstein, 1906, p. 632. In Einstein's notation,

$$\xi = \frac{\sum (m_\nu x_\nu) + \int x\rho_e d\tau}{\sum m_\nu + \int \rho_e d\tau}.$$

[47] Einstein, 1906, p. 633. In Einstein's notation,

$$(6) \quad \frac{d\xi}{dt} = \text{const.}$$

[48] "Es hat mich etwas befremdet, dass Sie bezüglich des Zusammenhanges zwischen träger Mass und Energie meine Priorität nicht anerkennen". Dukas Helen, and Hoffmann, Banesh, Albert Einstein, *The Human Side, New Glimpses from his Archives*, Princeton: Princeton University Press.1979. p. 20; p. 126.